\documentclass[
    ,final            
  ]
  {aipproc}

\layoutstyle{6x9}

\begin{document}

\title{Extensive population synthesis of isolated neutron stars with field
decay}

\classification{97.60.Gb; 97.60.Jd}
\keywords      {pulsars, neutron stars}

\author{S.B. Popov}{
  address={Sternberg Astronomical Institute, 119991, Moscow, Russia}
}

\author{P.A. Boldin}{
  address={MEPhI, 115409, Moscow, Russia}
}

\author{J.A. Miralles}{
  address={University of Alicante, 03080, Spain}
}

\author{J.A. Pons}{
  address={University of Alicante, 03080, Spain}
}

\author{B. Posselt}{
  address={Harvard-Smithsonian Center for Astrophysics, MA 02138 USA}
}

\begin{abstract}
We perform population synthesis studies of different types of neutron stars
(thermally emitting isolated neutron stars, normal radio pulsars, magnetars)
taking into account the magnetic field decay and using results from the most
recent advances in neutron star cooling theory. For the first time, we
confront our results with observations using {\it simultaneously} the Log N
-- Log S distribution for nearby isolated neutron stars, the Log N -- Log L
distribution for magnetars, and the distribution of radio pulsars in the $P$
-- $\dot P$ diagram. For this purpose, we fix a baseline neutron star model
(all microphysics input), and other relevant parameters to standard values
(velocity distribution, mass spectrum, etc.), only allowing to vary
the initial magnetic field strength. We find that our theoretical model is
consistent with all sets of data if the initial magnetic field distribution
function follows a log-normal law with 
$<\log \left(B_0/\left[G\right] \right) > \sim 13.25$ and
$\sigma_{\log B_0}\sim 0.6$. The typical scenario includes about 10\% of
neutron stars born as magnetars, significant magnetic field decay during the
first million years of a NS life (only about a factor of 2 for low field
neutron stars but more than an order of magnitude for magnetars), and a mass
distribution function dominated by low mass objects. This model explains
satisfactorily all known populations. Evolutionary links between different
subclasses may exist, although robust conclusions are not yet possible.

We apply the obtained field distribution and the model of decay to study
long-term evolution of neuton stars till the stage of accretion from the
interstellar medium. It is shown that though the subsonic propeller stage
can be relatively long, initially highly magnetized neutron stars ($B_0
> \sim 10^{13}$ G) reach the accretion regime within the Galactic lifetime if their
kick velocities are not too large. The fact that in previous studies made
$>$10 years ago, such objects were not considered results in a slight
increase of the Accretor fraction in comparison with earlier conclusions.
Most of the neutron stars similar to the Magnificent seven are expected to
become accreting from the interstellar medium after few billion years of
their evolution. They are the main predecestors of accreting isolated
neutron stars.
\end{abstract}

\maketitle


\section{Introduction}

Young neutron stars (NSs) appear as sources of different visible nature:
radio
pulsars (PSR), soft gamma-ray repeators (SGRs), anomalous X-ray pulsars
(AXPs), rotating radio transients (RRATs), central compact objects in
supernova remnants (CCOs), cooling close-by NSs dubbed the Magnificent seven
(M7).  Some NSs show several types of activities. Step by step we realize
that in many cases we cannot speak about   different subpopulations with
different evolutionary paths: sources evolve from one type to another.
One way to explain it is to consider the magnetic field evolution.

 On one hand, it is known that at least the activity of magnetars (SGRs and
AXPs) is due to the release of their magnetic field energy. On another hand,
for most of other types of NSs there is no direct evidence in favour of
significant role of the magnetic field decay. This controversial situation
is explained in the scenario studied by  Pons et al. (see, for example,
\citep{PMG2009} and references therein). In this contribution we briefly
summarize the results published in \cite{ppmbp2010, bp2010}, which are based
on this scenario of field evolution. We perform population synthesis of
several subpopulations of isolated NSs to derive a universal initial
magnetic field distribution, and to study the destiny of old NSs in this
scenario.



\section{The models}

 Here we follow the paper \cite{ppmbp2010}. We use three models to perform
population synthesis of close-by cooling NSs observed by ROSAT, population
synthesis of galactic magnetars, and, finally, population synthesis of
isolated normal PSRs. 

\subsection{Close-by cooling NSs}

 Here our calculations
 are based on the model described, for example, in \cite{posselt2008} and
references therein. NSs of different masses (we use eight values from 1.1 to
1.76 $M_{\odot}$) and different initian magnetic fields (we use seven
values from $3\, 10^{12}$~G to $3\, 10^{15}$~G) are evolved in the solar
vicinity (initial distance $<3$~kpc) till they are detectable by ROSAT above
$\sim 0.01$~cts/s from several tens of pc. Interstellar absorption is taken
into account. Results are calculated for the ROSAT PSCP and confronted with 
observations.  For normalization it is assumed that 270 NSs are born inside
3 kpc from the Sun in a Myr. The Gould belt contribution is taken into
account. Outside the Belt NSs are mainly born in known large OB
associations.

 Interested readers can look at the on-line version of the population
synthesis of isolated close-by cooling NSs:
http://www.astro.uni-jena.de/Net-PSICoNS/ (Boldin, Popov, Tetzlaff, in
press).

\subsection{Magnetars}

 We made a simple model to  calculate Log N -- Log L distribution for
galactic magnetars which is confronted with the data on known sources of
this kind, and with limits derived by \cite{muno2008}.  As we calculate
distribution in luminosity, not in flux, we
do not take into account interstellar absorption.

 To calculate luminosities we apply the same cooling curves with additional
heating due to magnetic field decay which are used for the population
synthesis of close-by isolated NSs. The same mass spectrum was applied. As
before we consider that the initial magnetic fields are uncorrelated with
masses. 

 In this modeling no Monte Carlo simulation is done. Instead,
we use complete cooling tracks of NSs with different masses and magnetic
fields to estimate the whole Galactic population of NSs with a given
luminosity. Absolute numbers are obtained by normalization to the total
birth rate of NSs. We use the Galactic NS formation rate equal to 1/30
yrs$^{-1}$.

\subsection{Radio pulsars}

We have performed Monte Carlo simulations to
generate a synthetic PSR population and confront our models with
observations. The methodology employed in the simulations closely follows
the work by \cite{fgkaspi2006} but some parameters of their model are
allowed to change according to the results of our previous calculations. 
The main goal of this modeling is to answer the following question: can we obtain
a synthetic PSR population compatible with the observed one and
consistent with our previous description for magnetars and close-by isolated
NSs? To this end, we start from the optimal population model
parameters obtained by \citet{fgkaspi2006} and modify only the initial
period and magnetic field distributions to account for the effect of
magnetic field decay consistent with our model.

To generate the PSR synthetic population we first choose the parameters of
the NS at birth closely following the model described by
\cite{fgkaspi2006}, which we
briefly summarize.
The place of birth is obtained according to the distribution
of their progenitors (massive Population I stars) which are mainly
populating the Galactic disk and more precisely its arms. The velocity at
birth is distributed
according the exponential distribution with a mean value of 380 km~s$^{-1}$.

The spin period of the star at
birth, $P_0$, is chosen from a normal distribution with a mean value of
$<P_0>$ and standard deviation
$\sigma_{P_0}$. Of course, only positive values are allowed. The initial
magnetic field at the magnetic pole is obtained from a log-normal
distribution
with mean value $<\log \left(B_0/{\rm \left[ G \right]}\right)>$ 
and standard deviation
$\sigma_{\log B_0}$ .

Once we have chosen the properties of the NS at birth we solve the
appropriate differential equations to obtain the position, period and
magnetic field at the present time. We use a smooth model for the Galactic
gravitational
potential. The period evolution is obtained by assuming
that the rotation energy losses are due to magnetic dipolar emission
(orthogonal rotators),
where the magnetic field is obtained from
our magneto-thermal evolutionary models.

At the end of the Monte Carlo simulation we end up with a synthetic
population of PSRs to be compared with a given observed sample. We use the
PSRs detected in the Parkes Multibeam Survey (PMBS) sample
\citep{Lyne2008} and, to limit the
contamination of our sample by recycled PSRs, we further ignore the
PSRs with $P < 30$ ms or $\dot P < 0$, and those in binary systems. With
this restrictions, our resulting sample contains 977 objects.
We use the parameters for detectability in the survey, radio luminosity and
beaming
given in \cite{fgkaspi2006}.

\section{Results of population synthesis for young neutron stars}

Our main result is the following: we are able to explain three different
populations of isolated NSs (close-by cooling NSs, magnetars, PSRs)
using the same set of initial distributions in the framework of decaying
magnetic field. We derived ``the best'' model for the initial magnetic field
distribution: it is the log-gaussian
distribution with  $<\log \left(B_0/{\rm \left[ G \right] } \right)>\sim 13.25$ and
$\sigma_{\log B_0}\sim 0.6$. 
Of course, real distribution can be more complicated, but we can put a
constraint on the fraction of magnetars. Our results show that it should be
below 20\%, 
most probably $\sim 10$\%.

\section{Old neutron stars}

 We use ``the best model'' for the initial magnetic field distribution to
study the destiny of old isolated NSs in the Galaxy. We perform a population
synthesis similar to \cite{p2000}, but with significant upgrade both in the
model and in the parameters.

 We show that due to a significant fraction of highly magnetized NSs (which
was neglected in earlier studies) many objects can reach the stage of
accretion from the interstellar medium. In the solar vicinity about 40\% of
isolated NSs can be accreting. 

\section{Conclusions}

 Here we presented some results which can be considered as one of the first
steps towards the ``grand unification in neutron stars''
\cite{kaspi2010}. We described several types of isolated NSs in the
framework of decaying magnetic field using the same initial distributions.
In addition, we demonstrated that due to relatively large fraction of
initially highly magnetized NSs, many old objects can start to accrete from
the interstellar medium.


\begin{theacknowledgments}
Participations of SP in the conference was supported by the RFBR grant
09-02-00032.
The work of SP and PB was supported by RFBR and Federal program for
scientific staff (02.740.11.0575).
\end{theacknowledgments}



\bibliographystyle{aipproc}   

\bibliography{popov}


\end{document}